\documentclass{llncs}

\usepackage{etex}
\usepackage{delarray}
\usepackage{amssymb}
\usepackage{amsmath}
\usepackage{graphicx}
\usepackage{subfig}
\usepackage[usenames,dvipsnames]{color}
\usepackage[all]{xy}

\usepackage{aicescover}

\newcommand{\symmruleTwoByTwo}[6]
{
  $\begin{aligned}
    \renewcommand{\arraystretch}{1.2}
      #1_{#2 \times #3} 
	  \rightarrow &
	  \left( 
	    \begin{array}{@{}c@{\,}|@{\,}c@{}} 
		  #1_{TL} & #1_{BL}^T \\\hline 
		  #1_{BL} & #1_{BR} 
		\end{array} 
	  \right) \\
      \small \textnormal{where } & #1_{#4} \textnormal{ is } #5 \times #6
  \end{aligned}$
}

\newcommand{\ruleTwoByTwo}[6]
{
  $\begin{aligned}
    \renewcommand{\arraystretch}{1.2}
      #1_{#2 \times #3} 
	  \rightarrow &
	  \left( 
	    \begin{array}{@{}c@{\,}|@{\,}c@{}} 
		  #1_{TL} & #1_{TR} \\\hline 
		  #1_{BL} & #1_{BR} 
		\end{array} 
	  \right) \\
      \small \textnormal{where } & #1_{#4} \textnormal{ is } #5 \times #6
  \end{aligned}$
}

\newcommand{\ruleTwoByOne}[6]
{
  $\begin{aligned}
    \renewcommand{\arraystretch}{1.2}
      #1_{#2 \times #3} 
	  \rightarrow &
	  \left( 
	    \begin{array}{@{}c@{}} 
		  #1_{T} \\\hline 
		  #1_{B}
		\end{array} 
	  \right) \\
      \small \textnormal{where } & #1_{#4} \textnormal{ is } #5 \times #6
  \end{aligned}$
}

\newcommand{\ruleOneByTwo}[6]
{
  $\begin{aligned}
    \renewcommand{\arraystretch}{1.2}
      #1_{#2 \times #3} 
	  \rightarrow &
	  \left( 
	    \begin{array}{@{}c@{\,}|@{\,}c@{}} 
		  #1_{L} & #1_{R}
		\end{array} 
	  \right) \\
      \small \textnormal{where } & #1_{#4} \textnormal{ is } #5 \times #6
  \end{aligned}$
}

\newcommand{\ruleOneByOne}[3]
{
  $\begin{aligned}
    \renewcommand{\arraystretch}{1.2}
      #1_{#2 \times #3} 
	  \rightarrow &
	  \left( 
	    \begin{array}{@{}c@{}} 
		  #1
		\end{array} 
	  \right) \\
      \small \textnormal{where } & #1 \textnormal{ is } #2 \times #3
  \end{aligned}$
}

\usepackage{float}
\floatstyle{boxed}
\newfloat{mybox}{htb}{lob}
\floatname{mybox}{Box}
\newsubfloat{mybox}

\usepackage{booktabs}
\usepackage{colortbl}

\definecolor{known}{RGB}{0, 100, 0} 
\definecolor{unknown}{rgb}{1.0, 0.0, 0.0} 
\newcommand{\known}[1]{\textcolor{known}{#1}}
\newcommand{\unknown}[1]{\textcolor{unknown}{#1}}
\newcommand{\myboxed}[1]{\boxed{\!\!{#1}\!\!}}

\newcommand{\click}{{\sc{Cl\makebox[.58\width][c]{1}ck}}}

\DeclareCaptionType{copyrightbox}

\newcommand{\prop}[2]{{\tt{#1(}}#2{\tt{)}}}

\begin{document}

\title{Knowledge-Based Automatic Generation\\of Partitioned Matrix Expressions}

\author{Diego Fabregat-Traver \and 
        Paolo Bientinesi}

\institute{
  AICES, RWTH Aachen, Germany \\
  \email{\{fabregat,pauldj\}@aices.rwth-aachen.de}
}

\aicescovertitle{Knowledge-Based\\Automatic Generation of\\Partitioned Matrix Expressions}
\aicescoverauthor{Diego Fabregat-Traver and Paolo Bientinesi} 
\aicescoverpublisher{\small{\bf * The final version published by Springer ({\url www.springerlink.com}) is available at:}\\
                    \url{http://link.springer.com/content/pdf/10.1007\%2F978-3-642-23568-9\_12.pdf}}
\aicescoverpage

\maketitle

\begin{abstract}

In a series of papers it has been shown that for many linear algebra
operations it is possible to generate families of algorithms by following
a systematic procedure.
Although powerful, such a methodology
involves complex algebraic manipulation, symbolic computations and
pattern matching, making the generation a process challenging to be
performed by hand.  We aim for a fully automated system that from the
sole description of a target operation creates multiple algorithms
without any human intervention.
Our approach consists of three main stages. The first stage yields the
core object for the entire process, the Partitioned Matrix Expression
(PME), which establishes how the target problem may be decomposed in
terms of simpler sub-problems.  In the second stage the PME is
inspected to identify predicates, the Loop-Invariants, to be used to
set up the skeleton of a family of proofs of correctness. In the third
and last stage the actual algorithms are constructed so that each of
them satisfies its corresponding proof of correctness.  In this paper
we focus on the first stage of the process, the automatic generation
of Partitioned Matrix Expressions. In particular, we discuss the steps
leading to a PME and the knowledge necessary for a symbolic system to
perform such steps. We also introduce \click{}, a prototype system
written in Mathematica that generates PMEs automatically.

\end{abstract}

\section{Introduction}

In the context of the Formal Linear Algebra Methods Environment
(FLAME) project~\cite{FLAMEOnline}, a methodology for the systematic
derivation of algorithms for matrix operations has been developed and
demonstrated. 
The approach has been successfully applied to all the 
operations included in the BLAS~\cite{level3BLAS} and 
RECSY~\cite{RECSY1,RECSY2} libraries and to
many included in the LAPACK~\cite{laug} library.
In general, the methodology
applies to any operation that can be expressed in a ``divide and
conquer'' fashion.
As opposed to the concept of ``Autotuning'', which
indicates the automatic tuning of a given
algorithm~\cite{atlas-sc98,FFTW05,Pueschel:05}, the word derivation refers to the
actual generation of both algorithms and routines to solve a given
target equation~\cite{Bientinesi:2005:SDD}.
The remarkable results achieved using this methodology are the 
subject of a series of publications. 
a) For many standard operations, e.g. the Cholesky and
the LU factorizations, all the previously known algorithms were systematically 
discovered, unifying them under a common root~\cite{FLAWN11}. 
b) For more involved operations like the Sylvester
equation and the reduction of a generalized eigenproblem to standard form,
the generated family of algorithms included new and better performing 
ones~\cite{Quintana-Orti:2003:FDA,FLAWN56}.
c) A related methodology for systematic analysis of round-off errors yielded bounds tighter than
those previously known~\cite{Paolo-MASA}.

\begin{figure}
\centering
    \includegraphics[scale=1]{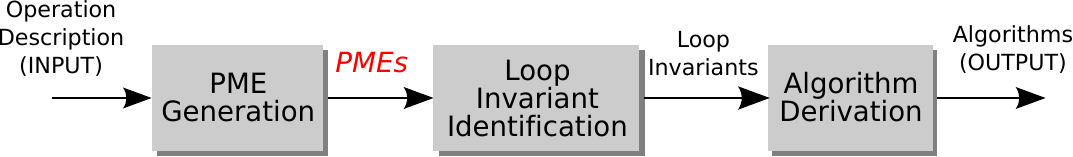}
    \caption{The three main stages in the process of derivation of algorithms.} \label{fig:steps}
\end{figure}

Although successful, the approach presents some limitations. The
algorithms are generated through complex symbolic computations, steps
often too complicated to be carried out by hand. Motivated by these
difficulties, we aim for a symbolic system that, given as input the
description of a matrix equation $Eq$, applies the steps dictated by
the FLAME methodology to derive a family of algorithms to solve $Eq$.
As shown in Fig.~\ref{fig:steps}, the procedure
consists of three successive stages---PME Generation, Loop-Invariant Identification, Algorithm Derivation---and is entirely determined by the mathematical description of the input operation.

In the first stage, the Partitioned Matrix Expression (PME) for the input 
operation is obtained. A PME is a decomposition of the original problem 
into simpler sub-problems in a ``divide and conquer'' fashion, exposing the
computation to be performed in each part of the output matrices. 
An example is shown in Box~\ref{box:PMELU}. The second 
stage of the process deals with the identification of Boolean predicates, 
the Loop-Invariants, that describe the intermediate state of computation 
for the sought-after algorithms. Loop-invariants can be extracted from the 
PME, and are at the heart of the automation of the third stage. In the third 
and last stage of the methodology, each loop-invariant is used to set up
a proof of correctness around which the algorithm is finally built.
Notice that the objective is not proving the correctness of a given
algorithm; vice-versa, the loop-invariant is chosen {\em before} the
algorithm is built. Indeed, the algorithm is constructed to 
satisfy a given proof of correctness, i.e., to possess the chosen 
loop-invariant.

\begin{mybox}
\vspace{2mm}
  \centering
	$
        \renewcommand{\arraystretch}{1.2}
		\left( {\begin{array}{@{}c@{}} 
			X_{T} = \Omega(L_{TL}, U, C_{T}) \\\hline
			X_{B} = \Omega(L_{BR}, U, C_{B} - L_{BL} X_{T}) 
		\end{array}} \right)
    $ 
\vspace{1mm}
	\caption{Partitioned Matrix Expression for the triangular Sylvester equation.} \label{box:PMELU}
\end{mybox}

\begin{figure}[t]
  \centering
  \includegraphics[scale=1.00]{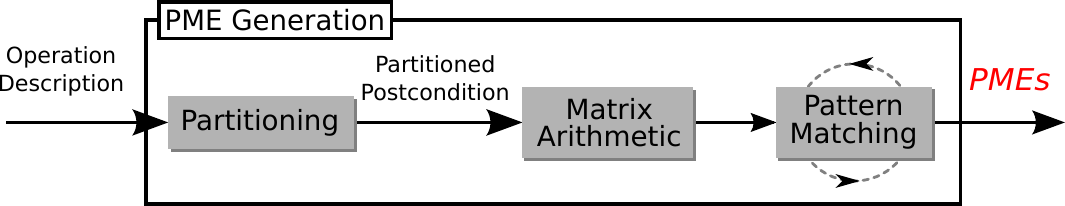}
  \caption{Steps for the automatic generation of PMEs.} \label{fig:stepsPME}
\end{figure}

This paper centers around the first stage of the derivation process, the generation of PMEs.
To this end we introduce a formalism to input into the
system the minimum amount of knowledge about the operation required by a system to 
perform all the subsequent stages automatically.
We then describe the process for transforming an input equation into PMEs.
As Fig.~\ref{fig:stepsPME} shows, such process involves three steps: 
1) the partitioning of the operands in the equation, 
2) matrix arithmetic involving the partitioned operands, and
3) a sequence of iterations, each consisting of algebraic manipulation and
pattern matching.
We demonstrate that the process can indeed be automated through
\click{}\footnote{The name \click{} epitomizes the idea 
that the effort a user has to make to obtain algorithms consists in just {\it one click}.}, 
a symbolic system written in Mathematica~\cite{MathematicaOnline} 
that performs all the steps for the PME generation.

The paper is organized as follows. In Sect.~\ref{sec:input} we categorize
the input needed by a symbolic system. Partitionings of the operands and 
inheritance of properties are discussed in Sect.~\ref{sec:partitioning}, 
while in Sect.~\ref{sec:PattMatch} we describe how to use partitionings 
to obtain PMEs. We draw conclusion in Sect.~\ref{sec:conclusions}.

\section{Input to the System} \label{sec:input}

Our first concern is to establish how a target operation should be
formally described.  
Since we are
aiming for a fully-automated system, i.e., without any human
intervention, we need a formalism to unequivocally describe a target
equation.
We choose the language traditionally used to reason about program correctness: 
equations shall be specified by means of the predicates Precondition ($P_{\rm pre}$) and
Postcondition ($P_{\rm post}$)~\cite{GrSc:92}. The precondition enumerates
the operands that appear in the equation and describes their properties, while
the postcondition specifies 
the equation to be solved.

The Cholesky factorization will serve as an example:
given a symmetric positive definite (SPD) matrix $A$, the goal is to find a
lower triangular matrix $L$ such that $L L^T = A$.
Box~\ref{box:CholOpDesc} contains the predicates $P_{\rm pre}$ and $P_{\rm post}$
relative to the Cholesky factorization; 
the notation $L = \Gamma(A)$ indicates that $L$ is the Cholesky factor of $A$.
\begin{mybox}
$$
\small
L = \Gamma(A) \equiv
\left\{
\begin{split}
P_{\rm pre}: \{ & \prop{Unknown}{L} \; \wedge \; \prop{LowerTriangular}{L} \;\; \wedge \\
            & \prop{Known}{A} \; \wedge \; \prop{SPD}{A} \} \\
\\[-4mm]
P_{\rm post}: \{ &  L L^T = A \}
\end{split}
\right.
$$
\caption{Formal description for the Cholesky factorization.}
\label{box:CholOpDesc}
\end{mybox}

Even though such a definition is unambiguous, it does not include all the
information needed by a symbolic system to fully automate the derivation
process. In Sect.~\ref{subsec:pattLearn} we discuss how a system 
expands its knowledge by ``learning of'' new equations, and in
Sect.~\ref{sec:partitioning} we overview the ground knowledge that a system must
possess relative to matrix partitioning and inheritance of properties.

\subsection{Pattern Learning}\label{subsec:pattLearn}

We refer to the pair of predicates ($P_{\rm pre}$ and $P_{\rm post}$)
in Box~\ref{box:CholOpDesc} as the {\em pattern} that identifies the
Cholesky factorization.  Such a pattern establishes that matrices $L$
and $A$ are one the Cholesky factor of the other provided that the
constraints in the precondition are satisfied, and $L$ and $A$ are
related as dictated in the postcondition ($L L^T = A$).  For instance,
in the expression
$$X X^T = A - B C,$$ 
in order to determine whether $X = \Gamma(A - B C)$, 
the following facts need to be asserted:
i) $X$ is an unknown lower triangular matrix; 
ii) the expression $A - B C$ is a known quantity ($A, B$ and $C$ are known); 
iii) the matrix $A - B C$ is symmetric positive definite.

The strategy for decomposing an equation in terms of simpler problems
greatly relies on pattern matching.  In the next section we describe
how matrix equations can be rewritten in terms of sub-matrices,
resulting in expressions seemingly more complicated than the initial
formulation. Such expressions are thus inspected to find segments 
corresponding to known patterns.

Initially, \click{} only knows the patterns for a basic set
of operations: addition, multiplication, inversion, and transposition
of matrices, vectors and scalars. This information is hard-coded.
More complex patterns are instead discovered during the process of PME
generation. For instance, the first time the PME for the Cholesky
factorization is generated, \click{} learns and stores the pattern
specified by Box~\ref{box:CholOpDesc}. Thanks to such patterns it will
then be possible to identify that a Cholesky factorization may be
decomposed into a combination of triangular systems and simpler
Cholesky factorizations. As \click{}'s pattern knowledge
increases, also does its capability of tackling complex operations.

\section{Partitioning and Inheritance} \label{sec:partitioning}

In this section we illustrate the first step towards the PME generation:
the partitioning of the operands (Fig.~\ref{fig:stepsPME}). 
To this end we introduce a set of rules to
partition and combine operands and to assert properties of expressions
involving sub-operands. The application of these rules to the
postcondition yields a predicate called {\em partitioned
postcondition}. Due to constraints imposed by both the structure of the input
operands and the postcondition, only few partitioning rules will be
admissible.

\subsection{Operands Partitioning and Direct Inheritance}

As shown in Box~\ref{box:part}, a generic matrix $A$ can be
partitioned in four different ways. The $1 \times 1$ rule
(Box~\ref{box:part}\subref{sbox:part1x1}) is special as it does not affect the operand;
we refer to it as the {\em identity}.
For a vector, only the $2 \times 1$ and $1 \times 1$ rules apply,
while for scalars only the identity is admissible.
When referring to any of the parts resulting from a non-identity rule, we use the
terms sub-matrix or sub-operand, and for $2 \times 2$ partitionings we 
also use the term quadrant.
 
\begin{mybox}
\vspace{1mm}
\begin{center}
    \subfloat[$2 \times 2$ rule]{
        \label{sbox:part2x2}
        \begin{minipage}{3.6cm}
		  \centering
		  \ruleTwoByTwo{A}{m}{n}{TL}{k_1}{k_2}
        \end{minipage}
    }
    \qquad
    \subfloat[$2 \times 1$ rule]
    {\label{sbox:part2x1}
        \begin{minipage}{3.6cm}
		  \centering
		  \ruleTwoByOne{A}{m}{n}{T}{k_1}{n}
        \end{minipage}
    }
    \\
    \subfloat[$1 \times 2$ rule]
    {\label{sbox:part1x2}
        \begin{minipage}{3.6cm}
		  \centering
		  \ruleOneByTwo{A}{m}{n}{L}{m}{k_2}
        \end{minipage}
    }
    \qquad
    \subfloat[$1 \times 1$ (identity) rule]
    {\label{sbox:part1x1}
        \begin{minipage}{3.6cm}
		  \centering
		  \ruleOneByOne{A}{m}{n}
        \end{minipage}
    }
\vspace{-3mm}
\end{center}
 
\caption{
  Rules for  partitioning  a generic matrix operand A.
  We use the subscript letters $T$, $B$, $L$, and $R$ for $T$op, $B$ottom,
  $L$eft, and $R$ight, respectively.
}\label{box:part}
\end{mybox}

The inheritance of
properties plays an important role in subsequent stages of the
algorithm generation process.  
Thus, when the operands have a special structure, it is beneficial to choose
partitioning rules that respect the structure.  
For a symmetric matrix, for instance,
it is convenient to create sub-matrices that exhibit the same property.
The $1 \times 2$ and $2 \times 1$ rules break the structure of a
symmetric matrix, as neither of the two sub-matrices inherit the
symmetry. Therefore, we only allow $1 \times 1$ or $2 \times 2$
partitionings, with the extra constraint that the $TL$ quadrant has to
be square.

Box~\ref{box:partLM} illustrates
the admissible partitionings for symmetric matrices.
On the left, the identity rule is applied and the operand remains unchanged.
On the right instead, a constrained $2 \times 2$ rule is applied, 
so that some of the resulting quadrants inherit properties.
Both $M_{TL}$ and $M_{BR}$ are square and symmetric,
and $M_{BL} = M_{TR}^T$ (or vice versa $M_{TR} = M_{BL}^T$).
Each matrix type allows specific partitioning rules and inheritance of properties; 
for triangular, diagonal, symmetric, and SPD matrices 
a library of admissible partitioning rules is incorporated into \click{}.

\begin{mybox}
\vspace{2mm}
\begin{center}
         \begin{tabular}{lcl}
        \begin{minipage}{3.3cm}
		  \centering
		  \ruleOneByOne{M}{m}{m}
        \end{minipage}
         \quad or \qquad &
        \begin{minipage}{3.6cm}
		  \centering
		  \symmruleTwoByTwo{M}{m}{m}{TL}{k}{k}
        \end{minipage}
         \end{tabular}\\[-3mm]
\caption{Partitioning rules for structured matrices.}\label{box:partLM}
\end{center}
\end{mybox}

\subsection{Theorem-aware Inheritance}\label{subsec:theorem-aware}

Although frequent, direct inheritance of properties is only the simplest form of inheritance.
Here we expose a more complex situation.
Let A be an SPD matrix. Because of symmetry, the only admissible partitioning rules are
the ones listed in Box~\ref{box:partLM}; applying the $2 \times 2$ rule, we obtain
\begin{equation}
\label{eqn:SPDPart}
	\begin{aligned}
    \renewcommand{\arraystretch}{1.2}
      A_{m \times m}  
	  \rightarrow &
	  \left(
	    \begin{array}{@{}c@{\,}|@{\,}c@{}} 
		  A_{TL} & A_{BL}^T \\\hline 
		  A_{BL} & A_{BR} 
		\end{array} 
	  \right) \\
      \small \textnormal{where } & A_{TL} \textnormal{ is } k \times k
  \end{aligned},
\end{equation}
and both $A_{TL}$ and $A_{BR}$ are symmetric. More properties about
the quadrants of $A$ can be stated. For example, it is well known that
{\it if $A$ is SPD, then every principal sub-matrix of $A$ is also
  SPD}.  As a consequence, the quadrants $A_{TL}$ and $A_{BR}$ inherit
the SPD property.  Moreover, it can be proved that given a $2 \times
2$ partitioning of an SPD matrix as in (\ref{eqn:SPDPart}), the
following matrices (known as Schur complements) are also symmetric
positive definite:\\[2mm]
\noindent
i) $\;A_{TL} - A_{BL}^{T} A_{BR}^{-1} A_{BL},$\\[1mm]
ii) $A_{BR} - A_{BL} A_{TL}^{-1} A_{BL}^{T}.$\\[-2mm]

The knowledge emerging from this theorem is hard-coded into \click{}.
In Sect.~\ref{sec:PattMatch} it will become apparent how this
information is essential for the generation of PMEs.

\subsection{Combining the Partitionings}

The admissible rules are now applied to rewrite the
postcondition.  Since in general each operand can be decomposed in
multiple ways, not one, but many partitioned postconditions are
created. As an example, in the Cholesky factorization
(Box~\ref{box:CholOpDesc}) both the $1\times 1$ and $2\times 2$ rules
are viable for both $L$ and $A$, leading to four different rewrite sets 
(Tab.~\ref{tab:partPostcond}).

\begin{table*}
\centering
\scriptsize
\begin{tabular}{c | c | c | c} \toprule
\renewcommand{\arraystretch}{1.4}
{\bf \#} & {\bf L} & {\bf A} & {\bf Partitioned Postcondition} \\ \midrule
\rowcolor[gray]{.9}
\rule[-0.3cm]{0cm}{0.7cm} 1 &
$
L \rightarrow \left( L \right)$ & $A \rightarrow \left( A \right)$ &
\renewcommand{\arraystretch}{1.4}
$
\left( L \right)
\left( L \right)^T
=
\left( A \right)
$ \\
\rule[-0.4cm]{0cm}{1cm} 2 &
$ L \rightarrow \left( L \right)$ &
\renewcommand{\arraystretch}{1.4}
$
         A \rightarrow
         \left( 
	    	\begin{array}{@{}c@{\,}|@{\,}c@{}} 
			  A_{TL} & A_{BL}^T \\\hline 
			  A_{BL} & A_{BR} 
			\end{array} 
		\right)
$ &
\renewcommand{\arraystretch}{1.4}
$
\left( L \right)
\left( L \right)^T
=
\left( 
  \begin{array}{@{}c@{\,}|@{\,}c@{}} 
    A_{TL} & A_{BL}^T \\\hline 
	A_{BL} & A_{BR} 
  \end{array} 
\right)
$ \\
\rowcolor[gray]{.9}
\rule[-0.4cm]{0cm}{1cm} 3 &
\renewcommand{\arraystretch}{1.4}
$
         L \rightarrow
         \left( 
           \begin{array}{@{}c@{\,}|@{\,}c@{}} 
		     L_{TL} & 0 \\\hline 
			 L_{BL} & L_{BR} 
		   \end{array} 
		 \right)
$ & $A \rightarrow \left( A \right)$ &
\renewcommand{\arraystretch}{1.4}
$
\left( \begin{array}{@{}c@{\,}|@{\,}c@{}} L_{TL} & 0 \\\hline L_{BL} & L_{BR} \end{array} \right)
\left( \begin{array}{@{}c@{\,}|@{\,}c@{}} L_{TL}^{T} & L_{BL}^{T} \\\hline 0 & L_{BR}^{T} \end{array} \right)
=
\left( A \right)
$ \\
\rule[-0.4cm]{0cm}{1cm} 4 &
\renewcommand{\arraystretch}{1.4}
$
         L \rightarrow
         \left( \begin{array}{@{}c@{\,}|@{\,}c@{}} L_{TL} & 0 \\\hline L_{BL} & L_{BR} \end{array} \right)
$ &
\renewcommand{\arraystretch}{1.4}
$
         A \rightarrow
         \left( \begin{array}{@{}c@{\,}|@{\,}c@{}} A_{TL} & A_{BL}^T \\\hline A_{BL} & A_{BR} \end{array} \right)
$ &
\renewcommand{\arraystretch}{1.4}
$
\left( \begin{array}{@{}c@{\,}|@{\,}c@{}} L_{TL} & 0 \\\hline L_{BL} & L_{BR} \end{array} \right)
\left( \begin{array}{@{}c@{\,}|@{\,}c@{}} L_{TL}^{T} & L_{BL}^{T} \\\hline 0 & L_{BR}^{T} \end{array} \right)
=
\left( \begin{array}{@{}c@{\,}|@{\,}c@{}} A_{TL} & A_{BL}^T \\\hline A_{BL} & A_{BR} \end{array} \right) $ \\
\bottomrule
\end{tabular}\\[1mm]
\caption{Application of the different combinations of partitioning rules to the postcondition.} \label{tab:partPostcond}
\end{table*}

It is apparent that some of the expressions in the
fourth column of Tab.~\ref{tab:partPostcond} are not algebraically well defined.  
The rules in the
second and third rows lead to ill-defined partitioned postconditions,
thus they should be discarded.
Despite leading to a well defined expression, the first row of the table should
be discarded too, as the goal is a {\it Partitioned} Matrix Expression
and it leads to an expression 
in which none of the operands has been partitioned.
In light of these additional restrictions, the only viable set
of rules for the Cholesky factorization is the one given in the last row of 
Tab.~\ref{tab:partPostcond}. 

In summary, partitioning rules must satisfy
both the constraints due to the nature of the individual operands, 
and those due to the operators appearing in the postcondition. 
In the next section we detail the algorithm used by \click{} to
generate only the viable sets of partitioning rules.

\subsection{Automation} \label{subsec:automation}

We show how \click{} performs the partitioning process automatically.
The naive approach would be to exhaustively search among all the
rules applied to all the operands, leading to a search space of
exponential size in the number of operands. 
Instead, \click{} utilizes an algorithm that traverses once the tree
that represents the postcondition in prefix notation and yields only
the viable sets of partitioning rules. 
The input to the algorithm is the predicates $P_{\rm pre}$ and $P_{\rm{post}}$ for a target operation.
As an example we look at the triangular Sylvester equation
$L X + X U = C,$
defined using our formalism as in Box~\ref{box:sylvdesc}.

\begin{mybox}
{\small
\begin{equation} \nonumber
X= \Omega(L, U, C) \equiv
\left\{
\begin{split}
P_{\rm pre}: \{ & \prop{Known}{L} \wedge \prop{LowerTriangular}{L} \, \wedge \\
            & \prop{Known}{U} \wedge \prop{UpperTriangular}{U} \, \wedge \\
            & \prop{Known}{C} \wedge \prop{Unknown}{X} \} \\
\\[-2mm]
P_{\rm post}: \{ &  L X + X U = C \}.
\end{split}
\right.
\end{equation}
}
\caption{Formal description for the triangular Sylvester equation.}
\label{box:sylvdesc}
\end{mybox}

\begin{figure}
\begin{center}
\includegraphics[scale=0.30]{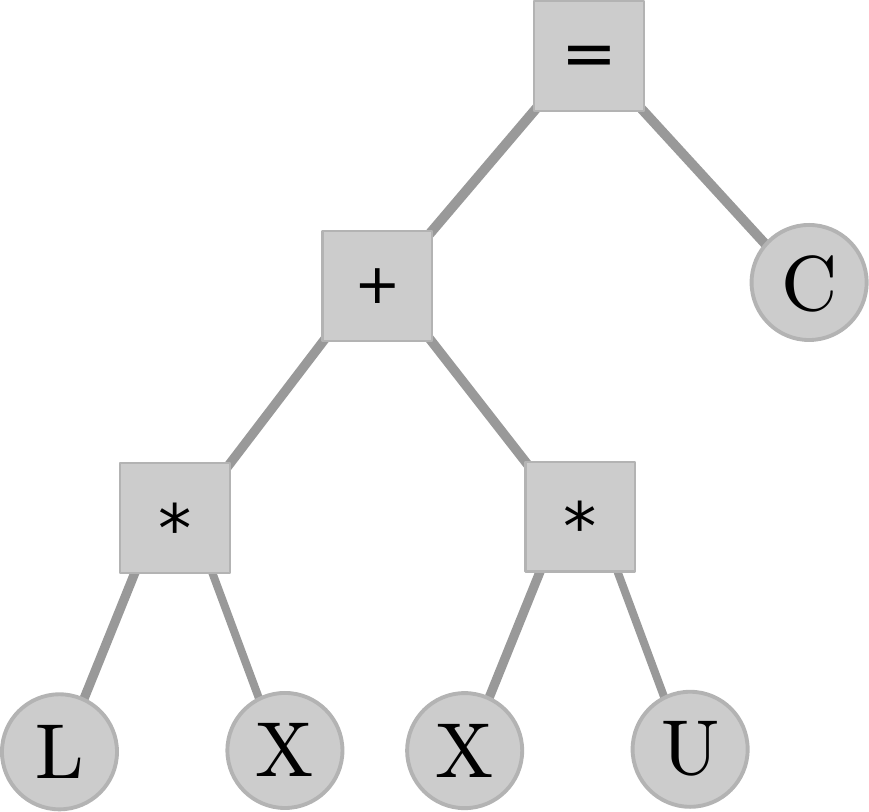}
\caption{Tree representation of the equation $L X + X U = C$.} \label{fig:tree}
\end{center}
\end{figure}

First, the algorithm transforms the postcondition to prefix notation
(Fig.~\ref{fig:tree}) and collects the name and the dimensionality of each operand. 
A list of disjoint sets, one per dimension of the operands is then
created.
This initial list for the Sylvester equation is
$\left[ \; \{ L_{r}\}, \{ L_{c}\}, \{ U_{r}\}, \{ U_{c}\}, \{ C_{r}\}, \{ C_{c}\}, \{ X_{r}\}, \{ X_{c}\} \; \right],$
where $r$ and $c$ stand for {\it rows} and {\it columns} respectively.
The algorithm traverses the tree, in a post-order fashion, to
determine if and which dimensions are bound together. Two dimensions
are bound to one another if the partitioning of one implies the
partitioning of the other. 
If two dimensions are found to be bound, then their corresponding
sets are merged together. As the algorithm moves from the leaves to
the root of the tree, it keeps track of the dimensions of the operands'
subtrees.

The algorithm starts by visiting the node corresponding to the operand $L$.
There it establishes that, since $L$ is lower triangular,
the identity and the $2 \times 2$ partitioning rules are the only admissible
ones. Thus, the rows and the columns of $L$ are bound together, and the list becomes
$\left[ \; \{ L_{r}, L_{c}\}, \{ U_{r}\}, \{ U_{c}\}, \{ C_{r}\}, \{ C_{c}\},\right.$
$\left.\{ X_{r}\}, \{ X_{c}\} \; \right].$
\noindent
The next node to be visited is that of the operand $X$.
Since $X$ has no specific structure, its analysis causes no bindings.
Then, the node corresponding to the $\ast$ operator is analyzed.
The dimensions of $L$ and $X$ have to agree according to the matrix product, therefore,
a binding between $L_c$ and $X_r$ is imposed:
$\left[ \; \{ L_{r}, L_{c}, X_{r}\}, \{ U_{r}\}, \{ U_{c}\}, \{ C_{r}\}, \{ C_{c}\}, \{ X_{c}\} \; \right].$
\noindent
At this stage the dimensions of the product $L X$ are also determined to be
$L_{r} \times X_{c}$.

The procedure continues by analyzing the subtree corresponding to the product $X U$.
Again, the lack of a specific structure in $X$ does not cause any binding and the algorithm follows
with the study of the node for the operand $U$. The triangularity of $U$ imposes a binding between
$U_{r}$ and $U_{c}$ leading to 
$\left[ \; \{ L_{r}, L_{c}, X_{r}\}, \{ U_{r}, U_{c}\}, \{ C_{r}\}, \{ C_{c}\}, \{ X_{c}\} \; \right].$
\noindent
Then, the node for the $\ast$ operator is analyzed, and a binding between $X_c$ and $U_r$ is found:
$\left[ \; \{ L_{r}, L_{c}, X_{r}\}, \{ U_{r}, U_{c}, X_{c}\},\right.$ $\left.\{ C_{r}\}, \{ C_{c}\} \; \right].$
The dimensions of the product $X U$ are determined to be $X_{r} \times U_{c}$. 

The next node to be considered is the corresponding to the $+$ operator. It imposes a binding between
the rows and the columns of the products $L X$ and $X U$, i.e., between $L_{r}$ and $X_{r}$,
and between $X_{c}$ and $U_{c}$. Since each of these pairs of dimensions already belong to the same
set, no modifications are made to the list. The algorithm establishes that the dimensions of 
the $+$ node are $L_r \times U_c$. Next, the node associated to the operand $C$
is analyzed. Since $C$ has no particular structure, its analysis does not cause any
modification. The last node to be processed is the equality operator
$=$. This node binds the rows of $C$ to those of $L$ ($C_r$, $L_r$)
and the columns of $C$ to those of $U$ ($C_c$, $U_c$). The final list
consists of two separate groups of dimensions:
$$\left[ \; \{ L_{r}, L_{c}, X_{r}, C_{r}\}, \{ U_{r}, U_{c}, X_{c}, C_{c}\} \; \right].$$

Having created $g$ groups of bound dimensions, 
the algorithm generates $2^g$ combinations of rules
(the dimensions within each group being either partitioned or not), 
resulting in a family of partioned postconditions, one per combination.
In practice, since the combination including solely identity rules
does not lead to a PME, only $2^g-1$ combinations are acceptable.
In our example the algorithm found two groups of bound dimensions, 
therefore three possible combinations of rules are generated:
1) only the dimensions in the second group are partitioned, 
2) only the dimensions in the first group are partitioned, or
3) all dimensions are partitioned.
The resulting partitionings are listed in Tab.~\ref{tab:sylvPart}.

\begin{table}
\centering
\scriptsize
\renewcommand{\arraystretch}{1.4}
\begin{tabular}{c | c | c | c | c} \toprule
{\bf \#} & {\bf L}  & {\bf U} & {\bf C} & {\bf X} \\\midrule
\rowcolor[gray]{.9}
\scriptsize 1 \rule[-0.35cm]{0cm}{0.9cm} &
$(L)$ & 
$\left( \begin{array}{@{}c@{\,}|@{\,}c@{}} U_{TL} & U_{TR} \\\hline 0 & U_{BR} \end{array} \right)$ &
$\left( \begin{array}{@{}c@{\,}|@{\,}c@{}} C_{L} & C_{R} \end{array} \right)$ &
$\left( \begin{array}{@{}c@{\,}|@{\,}c@{}} X_{L} & X_{R} \end{array} \right)$ \\
\scriptsize 2 \rule[-0.35cm]{0cm}{0.9cm} &
$\left( \begin{array}{@{}c@{\,}|@{\,}c@{}} L_{TL} & 0 \\\hline L_{BL} & L_{BR} \end{array} \right)$ &
$(U)$ &
$\left( \begin{array}{@{}c@{}} C_{T} \\\hline C_{B} \end{array} \right)$ &
$\left( \begin{array}{@{}c@{}} X_{T} \\\hline X_{B} \end{array} \right)$ \\
\rowcolor[gray]{.9}
\scriptsize 3 \rule[-0.35cm]{0cm}{0.9cm} &
$\left( \begin{array}{@{}c@{\,}|@{\,}c@{}} L_{TL} & 0 \\\hline L_{BL} & L_{BR} \end{array} \right)$ &
$\left( \begin{array}{@{}c@{\,}|@{\,}c@{}} U_{TL} & U_{TR} \\\hline 0 & U_{BR} \end{array} \right)$ &
$\left( \begin{array}{@{}c@{\,}|@{\,}c@{}} C_{TL} & C_{TR} \\\hline C_{BL} & C_{BR} \end{array} \right)$ &
$\left( \begin{array}{@{}c@{\,}|@{\,}c@{}} X_{TL} & X_{TR} \\\hline X_{BL} & X_{BR} \end{array} \right)$ \\\bottomrule
\end{tabular}\\[1mm]
\caption{Viable combinations of partitioning rules for the Sylvester equation.} \label{tab:sylvPart}
\end{table}

This very same process is used to find the bound dimensions of every target
operation and, accordingly, only each and every viable combination
of partitioning rules is generated.

\section{Matrix Arithmetic and Pattern Matching} \label{sec:PattMatch}

This section covers the second and third steps in the PME generation stage (Fig.~\ref{fig:stepsPME}).
Within the {\it Matrix Arithmetic} step, symbolic arithmetic is performed and 
the = operator is distributed over the partitions, 
originating multiple equations.
In (\ref{eqn:matArit1}) we display the result of these actions for the Cholesky factorization,
where the symbol $\star$ means that the equation in the top-right quadrant
is the transpose of the bottom-left one.

{\scriptsize
\vspace{-2mm}
\begin{eqnarray}
  \label{eqn:matArit1}
  \renewcommand{\arraystretch}{1.4}
  \left( \begin{array}{@{}c@{\,}|@{\,}c@{}} L_{TL} & 0 \\\hline L_{BL} & L_{BR} \end{array} \right) \!\!
  \left( \begin{array}{@{}c@{\,}|@{\,}c@{}} L_{TL}^{T} & L_{BL}^{T} \\\hline 0 & L_{BR}^{T} \end{array} \right)\!
  =\!
  \left( \begin{array}{@{}c@{\,}|@{\,}c@{}} A_{TL} & A_{BL}^T \\\hline A_{BL} & A_{BR} \end{array} \right)
  \Rightarrow 
  \notag 
  \renewcommand{\arraystretch}{1.4}
  \left( \begin{array}{@{}c@{\,}|@{\,}c@{}} L_{TL} L_{TL}^T = A_{TL} &
      \star \\\hline
      L_{BL} L_{TL}^T = A_{BL} &
      L_{BL} L_{BL}^T + L_{BR} L_{BR}^T = A_{BR}
          \end{array} \right) .  
\end{eqnarray}
} 

The {\em Pattern Matching} step delivers the sought-after PME.
Success of this process is dependent
on the ability to identify expressions with known structure and properties. 
In order to facilitate pattern matching, we force equations to be in
their {\em canonical form}. We state that an equation is in canonical
form if 
a) its left-hand side only consists of those terms that contain at least one unknown object,
and 
b) its right-hand side only consists of those terms that solely contain known objects.

This last step carries out an iterative process comprising three separate actions:
1) structural pattern matching:
equations are matched against known patterns;
2) once a known pattern is matched,
the unknown operands are flagged as known
and the equation becomes a tautology;
3) algebraic manipulation:
the remaining equations are rearranged in canonical form.
We clarify the iterative process by illustrating, action by action, how
\click{} works through the Cholesky factorization.
The first iteration is depicted in
Box~\ref{box:it1}, in
which
the top-left formula 
displays the initial state.
In all the next expressions, \known{green} and \unknown{red} are
used to highlight the known and unknown operands, respectively.

\paragraph{\bf Structural pattern matching:} 
All the equations in Box~\ref{box:it1}\subref{sbox:it1a} are in canonical form.
Through pattern matching, the top-left quadrant is the only one for which a match is found. 
\click{} identifies the equation as a Cholesky factorization (Box~\ref{box:it1}\subref{sbox:it1b}),
since the pattern in Box~\ref{box:CholOpDesc} is satisfied: the system recognizes that 
i) $L_{TL}$ is lower triangular;
ii) $A_{TL}$ is SPD; and 
iii) the structure of the equation matches the one in the postcondition ($L L^T = A$).

\paragraph{\bf Exposing new available operands:} 
Having matched the top-left equation, \click{}
turns the unknown operand \unknown{$L_{TL}$}
into \known{$L_{TL}$}, and propagates the information to all the other quadrants
(Box~\ref{box:it1}\subref{sbox:it1c}).
As a result, the top-left equation becomes a tautology.

\paragraph{\bf Algebraic manipulation:} 
All the remaining equations are still in canonical form,
 thus no operation takes place~(Box~\ref{box:it1}\subref{sbox:it1d}).

\begin{mybox} \centering
\tiny
\renewcommand{\arraystretch}{1.6}
\subfloat[Initial state.]
{ \label{sbox:it1a}
         $\left( \begin{array}{c|c}
                         \unknown{L_{TL}} \unknown{L_{TL}^T} = \known{A_{TL}} &
                            \star \\\hline
                            \unknown{L_{BL}} \unknown{L_{TL}^T} = \known{A_{BL}} &
                            \unknown{L_{BL}} \unknown{L_{BL}^T} + \unknown{L_{BR}} \unknown{L_{BR}^{T}} = \known{A_{BR}}
         \end{array} \right) $
}
\hspace{-.1cm}
\subfloat[Top-left equation is identified as a Cholesky sub-problem.]
{ 
\label{sbox:it1b}
         $\left( \begin{array}{c|c}
                         \unknown{\myboxed{L_{TL}}} = \Gamma(\known{A_{TL}}) &
                            \star \\\hline
                            \unknown{L_{BL}} \unknown{L_{TL}^T} = \known{A_{BL}} &
                            \unknown{L_{BL}} \unknown{L_{BL}^T} + \unknown{L_{BR}} \unknown{L_{BR}^{T}} = \known{A_{BR}}
         \end{array} \right) $
}
\\\vspace{1em}
\hspace{-.1cm}
\subfloat[$L_{TL}$ becomes a known operand for the rest of equations.]
{ 
  \begin{minipage}{5.4cm}
\label{sbox:it1c}
         $\left( \begin{array}{c|c}
                         \known{\myboxed{L_{TL}}} = \Gamma(\known{A_{TL}}) &
                            \star \\\hline
                            \unknown{L_{BL}} \known{\myboxed{L_{TL}^T}} = \known{A_{BL}} &
                            \unknown{L_{BL}} \unknown{L_{BL}^T} + \unknown{L_{BR}} \unknown{L_{BR}^{T}} = \known{A_{BR}}
         \end{array} \right) $
  \end{minipage}
}
\hspace{.50cm}
\subfloat[There is no need for algebraic manipulation.]
{ 
  \begin{minipage}{5.9cm}
\label{sbox:it1d}
         $\left( \begin{array}{c|c}
                         \known{L_{TL}} = \Gamma(\known{A_{TL}}) &
                            \star \\\hline
                            \unknown{L_{BL}} \known{L_{TL}^T} = \known{A_{BL}} &
                            \unknown{L_{BL}} \unknown{L_{BL}^T} + \unknown{L_{BR}} \unknown{L_{BR}^{T}} = \known{A_{BR}}
         \end{array} \right) $
  \end{minipage}
}
\caption{First iteration towards the PME generation.} \label{box:it1}
\end{mybox}

In this first iteration, one unknown operand, $L_{TL}$, has become known, 
and one equation has turned into a tautology.
The knowledge encoded in such a tautology is of importance for a subsequent iteration.
The {\bf second iteration} is shown in Box~\ref{box:it2}.

\paragraph{\bf Structural pattern matching:} 
Box~\ref{box:it2}\subref{sbox:it2a}
reproduces
the final state from the previous iteration.
Among the two outstanding equations,
the bottom-left one is identified (Box~\ref{box:it2}\subref{sbox:it2b}),
as it matches the pattern 
of a triangular system of equations with multiple right-hand sides ({\sc trsm}).
The pattern for a {\sc trsm} is
\begin{equation} \nonumber
\{
 X L^T = B \; \wedge \;
\prop{Output}{X} \; \wedge \; \prop{Input}{L} \; \wedge \;
\prop{LowerTriangular}{L} \; \wedge \; \prop{Input}{B} \}.
\end{equation}
For the sake of brevity,
we assume that \click{}
had learned such pattern from a previous derivation; 
in practice, in case the system does not know the pattern,
a nested task of PME generation would be initiated, 
yielding the required pattern.

\paragraph{\bf Exposing new available operands:} 
Once the {\sc trsm} is identified, the output operand $L_{BL}$ becomes
available and turns to green in the bottom-right quadrant
(Box~\ref{box:it2}\subref{sbox:it2c}).

\paragraph{\bf Algebraic manipulation:} 
The bottom-right equation is not in canonical form anymore:
the product $L_{BL} L_{BL}^T$, now a known quantity,
does not lay in the right-hand side. A simple manipulation
brings the equation back to canonical form (Box~\ref{box:it2}\subref{sbox:it2d}).

\begin{mybox} \centering
\tiny
\renewcommand{\arraystretch}{1.6}
\subfloat[Initial state.]
{ \label{sbox:it2a}
         $\left( \begin{array}{c|c}
                         \known{L_{TL}} = \Gamma(\known{A_{TL}}) &
                            \star \\\hline
                            \unknown{L_{BL}} \known{L_{TL}^T} = \known{A_{BL}} &
                            \unknown{L_{BL}} \unknown{L_{BL}^T} + \unknown{L_{BR}} \unknown{L_{BR}^{T}} = \known{A_{BR}}
         \end{array} \right) $
}
\hspace{-.2cm}
\subfloat[Bottom-left equation is identified as a triangular system of equations.]
{ \label{sbox:it2b}
         $\left( \begin{array}{c|c}
                         \known{L_{TL}} = \Gamma(\known{A_{TL}}) &
                            \star \\\hline
                            \unknown{\myboxed{L_{BL}}} = \known{A_{BL}} \known{L_{TL}^{-T}} &
                            \unknown{L_{BL}} \unknown{L_{BL}^T} + \unknown{L_{BR}} \unknown{L_{BR}^{T}} = \known{A_{BR}}
         \end{array} \right) $
}
\\\vspace{1em}
\hspace{-.1cm}
\subfloat[$L_{BL}$ becomes a known operand.] 
{ \label{sbox:it2c}
         $\left( \begin{array}{@{}c@{\,}|@{\,}c@{}}
                         \known{L_{TL}} = \Gamma(\known{A_{TL}}) &
                            \star \\\hline
                            \known{\myboxed{L_{BL}}} = \known{A_{BL}} \known{L_{TL}^{-T}} &
                            \known{\myboxed{L_{BL}^{\phantom{T}}}} \known{\myboxed{L_{BL}^T}} + \unknown{L_{BR}} \unknown{L_{BR}^{T}} = \known{A_{BR}}
         \end{array} \right) $
}
\subfloat[State after the algebraic manipulation.]
{ \label{sbox:it2d}
         $\left( \begin{array}{c|c}
                         \known{L_{TL}} = \Gamma(\known{A_{TL}}) &
                            \star \\\hline
                            \known{L_{BL}} = \known{A_{BL}} \known{L_{TL}^{-T}} &
                            \unknown{L_{BR}} \unknown{L_{BR}^{T}} = \known{A_{BR}} - \known{L_{BL}} \known{L_{BL}^T}
         \end{array} \right) $
}
\caption{Second iteration towards the PME generation.} \label{box:it2}
\end{mybox}

The process continues until all the equations are turned into tautologies.
The third and {\bf final iteration} for the Cholesky factorization
is shown in Box~\ref{box:it3}, where the top formula
replicates the final state from the previous iteration.

\paragraph{\bf Structural pattern matching:} 
Only one equation, the bottom-right one, remains unprocessed.  
At a first glance,
one might recognize a Cholesky factorization, but the corresponding
pattern in Box~\ref{box:CholOpDesc} requires $A$ to be SPD. The question is whether the expression $A_{BR} -
L_{BL} L_{BL}^T$ represents an SPD matrix. In
order to answer the question, \click{} applies rewrite rules and
symbolic simplifications.

In Sect.~\ref{subsec:theorem-aware} we explained
that the following quantities are known to be SPD: $A_{TL}$, $A_{BR}$,
$A_{TL} - A_{BL}^{T} A_{BR}^{-1} A_{BL}$, and $A_{BR} - A_{BL} A_{TL}^{-1} A_{BL}^{T}$.
In order to determine whether $A_{BR} - L_{BL} L_{BL}^T$
is equivalent to any of these expressions, 
\click{} makes use of the knowledge acquired throughout the previous iterations.
Specifically, in the first two iterations it was discovered that
$L_{TL} L_{TL}^T = A_{TL}$, and $L_{BL} = A_{BL} L_{TL}^{-T}$.
Using these tautologies as rewrite rules, the expression $A_{BR} -
L_{BL} L_{BL}^T$ is manipulated.  First, the equality $L_{BL} =
A_{BL} L_{TL}^{-T}$ is used to replace the instances of $L_{BL}$,
yielding $A_{BR} - A_{BL} L_{TL}^{-T} L_{TL}^{-1} A_{BL}^T$, and
equivalently, $A_{BR} - A_{BL} (L_{TL} L_{TL}^{T})^{-1}
A_{BL}^T$. Then, by virtue of the tautology $L_{TL} L_{TL}^T = A_{TL}$,
$L_{TL} L_{TL}^T$ is replaced by $A_{TL}$, yielding $A_{BR} - A_{BL}
A_{TL}^{-1} A_{BL}^{T}$. Now, this expression is known to be SPD.
Thanks to these manipulations, \click{} successfully associates the
bottom right equation with the pattern for a Cholesky factorization.

\paragraph{\bf Exposing new available operands:} 
Once the expression in the bottom-right quadrant is identified, 
the system exposes the quantity $L_{BR}$ as known.
Since no equation is left, the process completes and the PME---formed by the three
tautologies---is returned as output. 

\begin{mybox} \centering
\tiny
\renewcommand{\arraystretch}{1.6}
\subfloat[Initial state.]
{ \label{sbox:it3a}
         $\left( \begin{array}{c|c}
                         \known{L_{TL}} = \Gamma(\known{A_{TL}}) &
                            \star \\\hline
                            \known{L_{BL}} = \known{A_{BL}} \known{L_{TL}^{-T}} &
                            \unknown{L_{BR}} \unknown{L_{BR}^{T}} = \known{A_{BR}} - \known{L_{BL}} \known{L_{BL}^T}
         \end{array} \right) $
}
\hspace{-.1cm}
\subfloat[Bottom-right equation is identified as a Cholesky factorization.]
{ \label{sbox:it3b}
         $\left( \begin{array}{c|c}
                         \known{L_{TL}} = \Gamma(\known{A_{TL}}) &
                            \star \\\hline
                            \known{L_{BL}} = \known{A_{BL}} \known{L_{TL}^{-T}} &
                            \unknown{\myboxed{L_{BR}}} = \Gamma(\known{A_{BR}} - \known{L_{BL}} \known{L_{BL}^T})
         \end{array} \right) $
}
\\\vspace{-.5em}
\hspace{-.1cm}
\subfloat[$L_{BR}$ becomes a known operand.]
{ \label{sbox:it3c}
         $\left( \begin{array}{c|c}
                         \known{L_{TL}} = \Gamma(\known{A_{TL}}) &
                            \star \\\hline
                            \known{L_{BL}} = \known{A_{BL}} \known{L_{TL}^{-T}} &
                            \known{\myboxed{L_{BR}}} = \Gamma(\known{A_{BR}} - \known{L_{BL}} \known{L_{BL}^T})
         \end{array} \right) $
}
\subfloat[Final PME.]
{ \label{sbox:it3d}
         $\left(
         \begin{array}{c|c}
                 \known{L_{TL}} = \Gamma(\known{A_{TL}}) &
                 \star \\\hline
                 \known{L_{BL}} = \known{A_{BL}} \known{L_{TL}^{-T}} &
                 \known{L_{BR}} = \Gamma(\known{A_{BR}} - \known{L_{BL}} \known{L_{BL}^T})
         \end{array}
         \right) $
}
\caption{Final iteration towards the PME generation.} \label{box:it3}
\end{mybox}

By means of the described process, PMEs for a target equation are automatically
generated. The PME for the Cholesky factorization is given in Box~\ref{box:PMEChol}.

\begin{mybox} \centering
\vspace{-3mm}
	\renewcommand{\arraystretch}{1.4}
	$$\left( {\begin{array}{@{\,}c@{\,}|@{\,}c@{\,}} 
			L_{TL} = \Gamma(A_{TL}) &
			\star \\\hline
			L_{BL} = A_{BL} L_{TL}^{-T} &
			L_{BR} = \Gamma(A_{BR} - L_{BL} L_{BL}^T)
	 \end{array}} \right)$$ \\
\vspace{-3mm}
	\caption{Partitioned Matrix Expression for the Cholesky factorization.} \label{box:PMEChol}
\end{mybox}

\section{Conclusions} \label{sec:conclusions}

The work we presented sets the ground for the development of a
symbolic system that, from the sole description of an operation,
generates algorithms automatically. The core of our methodology stands
in the PME.  A PME encapsulates the information about the target
operation in a way that facilitates the subsequent identification of
loop-invariants. The loop-invariants then lead to the final algorithms
through a technique based on program correctness. In this paper we
introduce a symbolic system, \click{}, that automates the generation of
PMEs.

In order to generate PMEs, \click{} first identifies how the op\-er\-ands in the
operation may be partitioned. Instead of a brute force approach of
exponential complexity, \click{} utilizes a tree-based algorithm that
yields only the viable sets of partitioning rules.
Through a process of pattern matching, each such set leads to a distinct PME. 
The key in the PME generation is \click{}'s ability to identify known patterns. 
Initially, \click{} only recognizes elementary structures, but its
knowledge expands by automatically learning the patterns associated
with the operations it tackles. Thanks to this augmenting
internal knowledge, the system may generate PMEs for increasingly
complex operations.

To illustrate \click{}, we discussed the
Cholesky factorization and, partially (due to space constraints),
the Sylvester equation. Despite the fact
that such operations differ in multiple ways---number and properties
of the operands, number of valid sets of partitioning rules, number of
PMEs---the steps performed by \click{} leading to the PMEs are
exactly the same. As future work, we plan to add support
for higher dimensional objects
and the derivative operator.

\section{Acknowledgements}
The authors gratefully acknowledge the support received from the
Deutsche Forschungsgemeinschaft (German Research Association) through
grant GSC 111.


\end{document}